# Ferromagnetic structures in $Mn_2CoGa$ and $Mn_2CoAl$ doped by Co, Cu, V, and Ti

Y. J. Zhang(张玉洁), G. J. Li(李贵江), E. K. Liu(刘恩克), J. L. Chen(陈京兰), W. H. Wang(王文洪), G. H. Wu(吴光恒)[*]

*Beijing National Laboratory for Condensed Matter Physics, Institute of Physics, Chinese Academy of Sciences, Beijing 100190, People's Republic of China*

**Abstract:**

The structure and magnetic properties in doped Heusler alloys of $Mn_2CoGa$ and $Mn_2CoAl$ have been investigated by experiments and calculations. The main group elements of Ga and Al are substituted by the magnetic or non-magnetic transition metals, Co, Cu, V, and Ti in the alloy systems. Three kinds of local ferromagnetic structures, Co-Mn-Co, Mn-Co-Mn and Mn-Co-V, have been found. They embed in the native ferrimagnetic matrix and increase the magnetization with different increments. The Co-Mn-Co ferromagnetic structure shows the largest increment of $6.18\mu_B$ /atom. In addition, interesting results for non-magnetic Cu increasing the magnetization and the V atom having a large ferromagnetic moment of about $1.0\mu_B$ have been obtained. The exchange interaction energy can be increased by the newly added Co and depleted by supporting a ferromagnetic coupling in other substitution cases, and showing the variation of the $T_C$. Our calculation of electronic structure verifies the strong *d-d* hybridization when the three ferromagnetic structures are achieved. It has also been found that the covalent effect from the Ga and Al determines the generation of the local ferromagnetic structure and the tolerance for dopant content.

---

[*] Author to whom correspondence should be addressed. Electronic mail: ghwu@aphy.iphy.ac.cn.

## Ⅰ. INTRODUCTION

The Heusler alloys have been the significant representatives in ternary alloys.[1,2] Materials that exhibit many excellent properties, such as half-metallic magnetism,[3,4] magnetic-field-induced transformation,[5] and magnetocaloric effect,[6] are believed to be potential candidates for applications. Among them, alloys with Mn element have been widely investigated, because the magnetic moments of Mn atoms can be coupled in the style of the ferromagnetic or antiferromagnetic depending on their neighboring environment.[7-13] In these high-ordered intermetallic compounds, the magnetic atoms may occupy four non-equivalent crystallographic positions[1,14] if the elements of main group are substituted by the magnetic ones off-stoichiometrically. Therefore, in the case of existence of strong exchange interaction effect, the different atomic configuration can evolve into the various magnetic structures embedded each other[9].

The parent alloys used in previous works[9] adopted the $Mn_2NiGa$ and the Co was introduced to substitute the Ga. In these systems, the doped Co takes two roles, generating a local Mn-Co-Mn structure in the nearest neighboring distance and providing a strong exchange interaction. By using this parent alloy, however, it makes the Co irreplaceable and obstructs the doping with other elements, especially in the case of low composition tolerance. Therefore, the works for doping other magnetic elements, Fe, V, Ni, Cr, become sparse.

In the present work, Heusler alloys $Mn_2CoGa$ and $Mn_2CoAl$ with $Hg_2CuTi$ structure are chosen as the parent alloys, in which the strong exchange interaction has already been achieved due to the existence of native Co. The element substitution is performed by replacing the main group elements of Ga and Al to transition metals of Ti, V, Co and Cu. This enables ferromagnetic structures to be formed different from those in $Mn_2NiGa$.[9] Based on this strategy, three different ferromagnetic structures can be achieved in the native ferrimagnetic matrix of the parent alloys, which increases the magnetization with the various increments. The electronic structure calculation reveals the respective contribution of the magnetic atoms and indicates the *d-d* hybridization effect which forms these ferromagnetic structures. The exchange interaction effect has been investigated based on the experimental thermal-magnetic measurements. The covalence effect from the Ga and Al on the atomic configuration and the tolerance for dopant content has also been studied.

The paper is organized as follows: Section Ⅱ contains details concerning the methods of the experiment and the computation. Section Ⅲ generally discusses the basic structure and the atom occupation rule in Heusler alloys. Section IV presents the experimental and calculated results and related discussion, including: (A) the crystal structure; (B) the magnetization; (C) magnetic structures and calculated magnetic moments; (D) the exchange interaction and electronic structure calculation. Finally, the paper is summarized in Sec. V.

## II. EXPERIMENTAL AND COMPUTATIONAL DETAILS

Based on two parent alloys of $Mn_2CoGa$ and $Mn_2CoAl$, two series of polycrystalline samples $Mn_2CoM_xGa_{1-x}$ (M = Ti, V, Co, and Cu; $x = 0.0 \sim 0.4$) and $Mn_2Co_{1+x}Al_{1-x}$ ($x = 0.0 \sim 0.48$) were prepared by arc-melting in argon atmosphere, in which, the purity of elemental metals was higher than 99.95 %. The alloy ingots were heat-treated at 920 $^o$C for 24 hrs to homogenize the composition. In order to eliminate the second phase with faced-centered cubic (*fcc*) structure, the melt-spinning technique was adopted[15] for some samples with high concentration dopants. In order to ensure high chemical ordering, the homogenized and spun samples were further annealed at 650 $^o$C for 72 hrs, and subsequently quenched in an ice-water mixture.

Structural analyzes were carried out by x-ray diffraction (XRD) technique with Cu-Kα radiation

(λ=1.5418 Å) from the ground powder. The magnetic measurements were performed using a SQUID magnetometer (SQUID-Quantum Design). The virtual crystal approximation method was used to simulate a standard of atomic ordering x-ray diffraction pattern for comparing the experimental XRD profiles.

The electronic structure and magnetic properties were calculated using the Korringa–Kohn–Rostoker method combined with the coherent potential approximation and the local density approximation (KKR-CPA-LDA method).[16-18] This method is high speed, high precision and is a powerful method for disordered systems.[17, 19, 20]

**III. BASIC STRUCTURE AND OCCUPATION RULE OF HEUSLER ALLOYS**

The structure of Heusler alloy can be considered as four interpenetrating *fcc* lattices, which has four unique crystal sites,[1, 3] defined as *A* (0, 0, 0), *B* (0.25, 0.25, 0.25), *C* (0.5, 0.5, 0.5) and *D* (0.75, 0.75, 0.75) as shown in Fig. 1. Chemically, this structure is characterized by the formula $X_2YZ$, where X and Y stand for transition metals and Z represents main group element. In Heusler alloy, the main group element Z occupies *D* site; X and Y are distributed in *A*, *B* and *C* sites based on an empirical rule.[1, 21, 22] The atoms will preferentially occupy the *A* and *C* sites if they have a relatively larger number of valence electrons, while those with a relatively smaller number of valence electrons will preferentially occupy the *B* and *D* sites.[13, 23] Thus, there are two types of structures formed by the different sitting of X and Y elements. When X occupies the *A/C* site and Y occupies the *B* site, the structure is $L2_1$ type; while a $Hg_2CuTi$ structure will be formed if Y enters *C* site and X prefers *A* and *B* sites.[24] The Heusler alloys $Mn_2CoGa$ and $Mn_2CoAl$ used in the present work as the parent alloy are with the $Hg_2CuTi$ structure. Based on the previous studies about the intermetallic compounds,[25, 26] this atomic ordering phenomenon was attributed to the covalent effect caused by the *p-d* orbital hybridization between the main-group and the transition-metal atoms.[27-29]

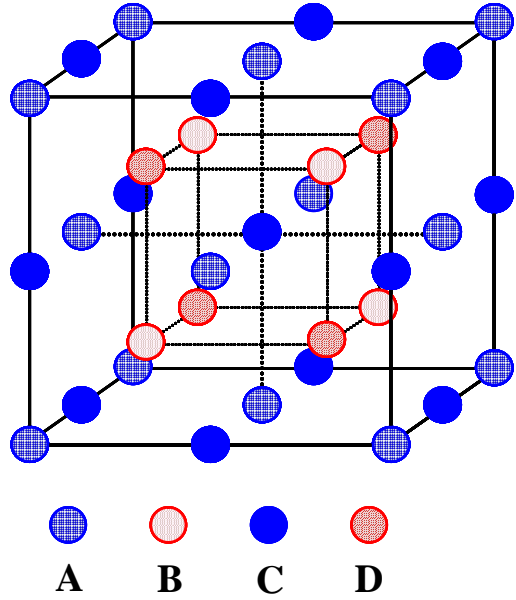

FIG. 1. (Color online) Atomic configuration of Heusler alloys.

## Ⅳ RESULTS AND DISCUSSIONS

### A. Crystal structure

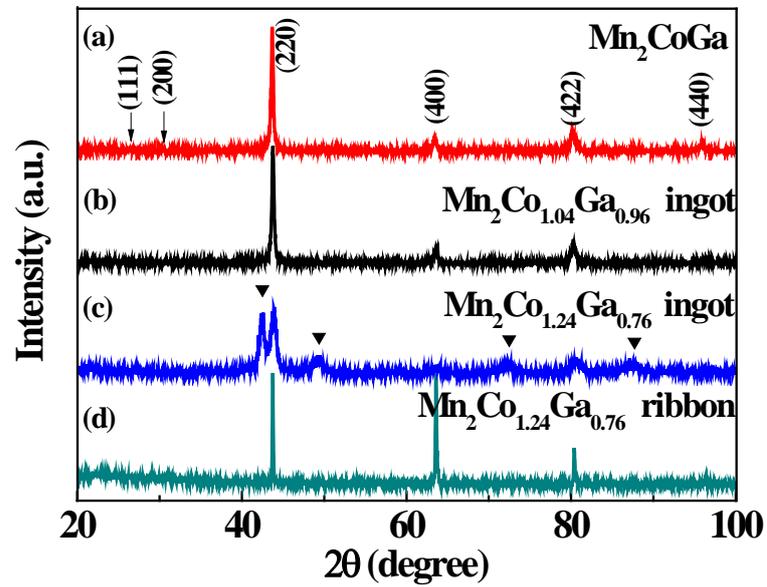

FIG. 2. (Color online) XRD patterns of $Mn_2Co_{1+x}Ga_{1-x}$ samples prepared by arc-melting (a~c) and melt-spinning (d).

Figure 2 shows the XRD patterns of some typical samples for $Mn_2Co_{1+x}Ga_{1-x}$ alloys. $Mn_2CoGa$ shows a typical body-centered cubic (*bcc*) structure. In the doping case, the arc-melting $Mn_2Co_{1+x}Ga_{1-x}$ samples persist in a pure phase until the $x$ is increased to 0.12. Fig. 2(c) shows the obvious *fcc* second phase, as marked, in the sample with $x = 0.24$. This is due to that the decreased of Ga content weakens the covalent bonding effect[25] from the main group elements and decreases the stability of the *bcc* phase. This second phase has low magnetization and changes the target component, so it should be eliminated by the melt-spinning method as shown in Fig. 2(d). Because the rapid solidification can increase the solubility above the equilibrium and refine the microstructure, the maximum Co doping concentration has been extended to $x = 0.40$ to obtain the pure *bcc* phase.[30] XRD patterns of all samples have been indexed to obtain the lattice constant for the series of samples, which are linearly fit and used in the calculation.

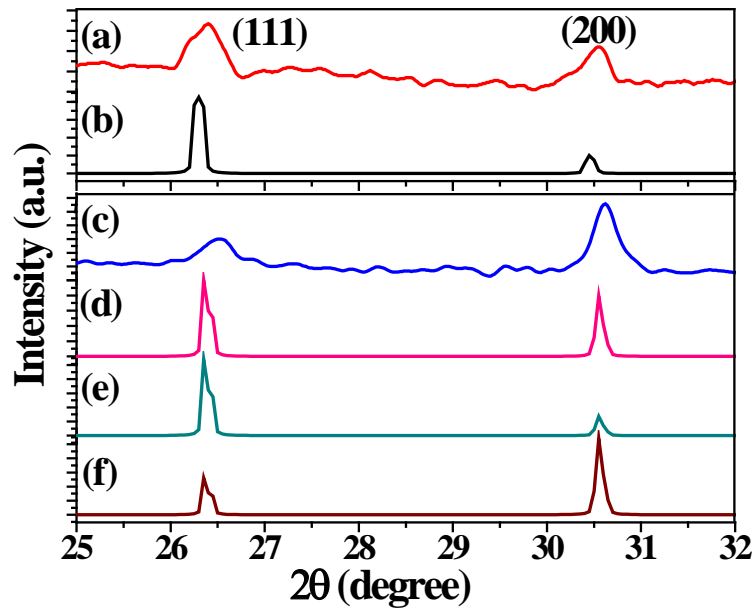

FIG. 3. (Color online) Experimentally measured superlattice peaks of (111) and (200) of $Mn_2CoGa$ (a) and $Mn_2CoAl$ (c) alloys. For comparing, the simulated superlattice peaks belong to $Mn_2CoGa$ state at $Hg_2CuTi$ (b) and $Mn_2CoAl$ stated at $Hg_2CuTi$ (d), $L2_1$ (e) and B2 (f) structures are also illustrated in the Figure.

The ratio of peak intensity of $(111)/(200)$[1,13] and $(200)/(220)$[31] is usually investigated to understand the degree of atomic ordering configuration in Heusler alloys. Fig. 3 (a) presents the (111) and (200) peaks of stoichiometric $Mn_2CoGa$. It can be found that the intensity ratio between (111) and (200) is consistent with the calculated results (Fig. 3(b)), which confirms that the $Mn_2CoGa$ states at the $Hg_2CuTi$ structure. However, a kind of disordered structure, B2-type structure (including A/C or B/D disordering) may occur and cause an abnormal (111)/(200) ratio. Turning to our $Mn_2CoAl$ alloy, it has relatively high (200) peak as shown in Fig. 3(c). Comparing with the calculated profiles (Fig. 3(d-e)), it belongs to neither $Hg_2CuTi$ nor $L2_1$ structures. Simulating some possible structures with relatively low ordered level, we find that the partly disordered (*B* and *D* sites occupied randomly, B2) structure has such ratio style. Further calculations indicate that, based on the experimental ratio, it corresponds to the case about 20% Mn and Ga disordered between the *B* and *D* sites in $Mn_2CoAl$ alloy(Fig. 3(f)). This situation was observed in $L2_1$-type Heusler alloys using Al as the main group element.[14, 32] Our work reveals that it can also occur in $Hg_2CuTi$-type Heusler alloys.

In the present work, the B2-type structure was observed more often in the materials with off-stoichiometrical composition. This is due to that the covalent effect is weakened by reducing the content of main group elements. However, it should be emphasized that such kind of B2 structure does not change the nearest neighboring relationship between the magnetic atoms and will not affect the magnetic properties as discovered in this work.

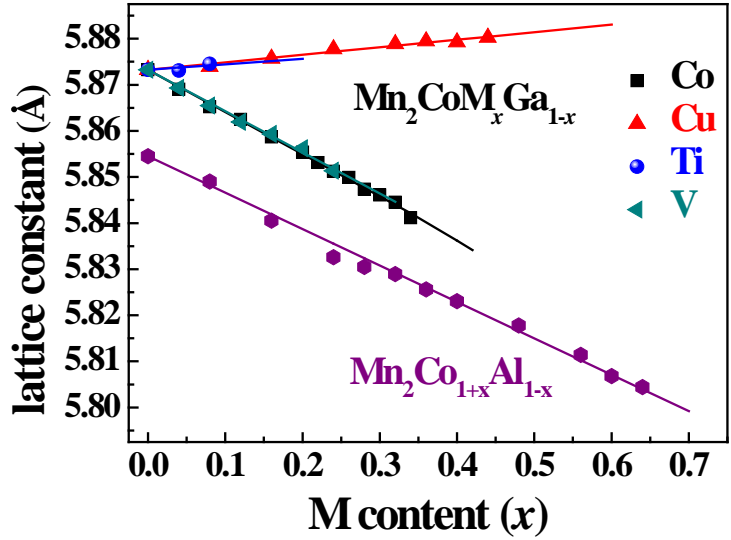

FIG. 4. (Color online) Lattice constants as a function of doping concentration $x$ in $Mn_2CoM_xGa_{1-x}$ (M = Co, Cu, V, and Ti) and $Mn_2Co_{1+x}Al_{1-x}$ alloys. The straight lines represent the linear fit.

Figure 4 displays the lattice constants as a function of doping concentration $x$ of Co, Cu, Ti and V in $Mn_2CoGa$. The result of $Mn_2Co_{1+x}Al_{1-x}$ is also illustrated in the Figure. It can be seen that all samples shows a linear change behavior. Stoichiometric $Mn_2CoGa$ has a lattice constant of $a = b = c = 5.873$ Å which is larger than that of $Mn_2CoAl$ (5.854 Å). It is usually attributed to the atomic size: the atomic radius of Ga (1.30 Å) is larger than that of Al (1.25 Å).[33] However, it is another case when the transition metals are doped in the parent alloy. Substituting the non-magnetic Cu and Ti for Ga in $Mn_2CoGa$, the lattices are expanded, while the magnetic Co and V doping causes the lattices shrink. This result seems to conflict with the dominating effect of atomic radius observed in many other alloys.[7] It suggests that the radii of doping atoms may be not the unique factor to determine the lattice constants. On the other hand, from Fig. 4, one may find that much more Co can be doped into $Mn_2CoAl$ than $Mn_2CoGa$, which may be due to the stronger covalent effect of Al than Ga.[25]

**B. Magnetization**

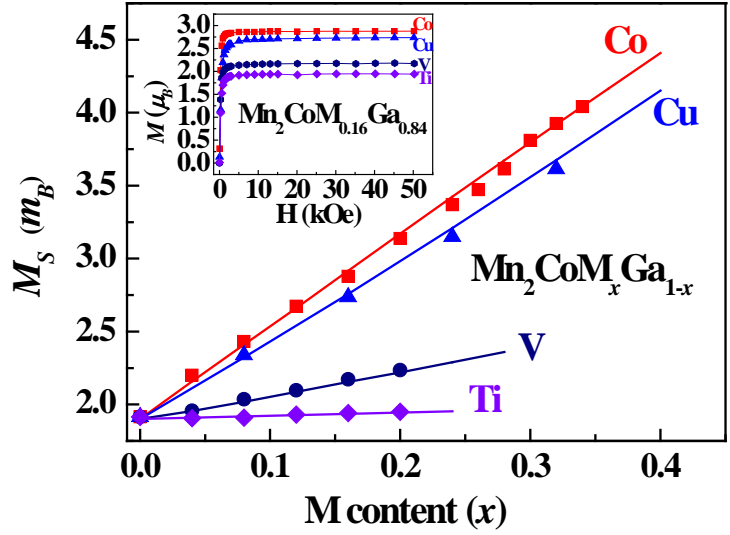

FIG. 5. (Color online) Concentration dependence of magnetic molecular moments in $Mn_2CoM_xGa_{1-x}$ (M = Co, Cu, V and Ti). The straight lines are the calculated results. The inset shows the magnetization curves of $Mn_2CoM_{0.16}Ga_{0.84}$ with four kinds of dopants measured at 5K.

Figure 5 shows the concentration dependence of the magnetic molecular moments in $Mn_2CoM_xGa_{1-x}$ samples. The moment values come from the saturation magnetization ($M_s$) measured at 5K. Additionally, the values calculated using the KKR-CPA-LDA method are listed in Table I and are also illustrated in Fig. 5 as the straight lines for comparison. The experimental results are quite consistent with the calculated ones, which prove the KKR-CPA-LDA method used in the present work is credible. One can see that the Co and Cu doping can increase the molecular moments rapidly, while V doping shows a small increment and Ti brings almost no increase, respectively. From the slope of the curves, one may find the molecular moment increments for each doped atom, $\Delta m$ are quite different.

Co doping leads to an increment of $\Delta m_{Co} = 6.18\mu_B$. Apparently, the moment of added Co atoms selves can not provide so much additional moment for the $Mn_2CoM_xGa_{1-x}$ alloy. Especially, we have found that there is a quite large increment of $\Delta m_{Cu} = 5.24\mu_B$ when the non-magnetic Cu doped in the system. On the other hand, as the same non-magnetic element of Ti, the increment of $\Delta m_{Ti} = 0.16\mu_B$ is near zero. For V doping, the increment of $\Delta m_V$ is $1.57\mu_B$, which implies the V atom may contribute some ferromagnetic moment. These results indicate there must be different variation of the magnetic structure occurred when the $Mn_2CoGa$ was doped with different elements. The related physical mechanism will be discussed associating with the calculation results.

**C. Magnetic structure**

In theoretical calculations, the $Hg_2CuTi$ structure is used for the parent alloys of $Mn_2CoGa$ and $Mn_2CoAl$. The lattice constants are taken by linear fitting the experimental results shown in Fig. 4. We regulate the site of doping atoms depending upon the occupation rule mentioned before. Thus, the dopants of the Ti and V occupy the $D$ site, while the Co and Cu occupy the Mn($A$) site with forcing the corresponding Mn($A$) atom to occupy the vacant $D$ site, although all the dopants chemically substitute Ga.

Table I. The calculated and experimentally measured molecular moments, $M_{Cal.}$ and $M_S$ ($\mu_B$ /f.u.), and the calculated atomic moments ($\mu_B$ /atom) for $Mn_2CoM_xGa_{1-x}$ alloys.

| Compounds | $M_S$ | $M_{Cal.}$ | Mn($A$) | Mn($B$) | Co($C$) | Ga | Mn($D$) | M($A$) | M($D$) |
|---|---|---|---|---|---|---|---|---|---|
| $Mn_2CoGa$ | 1.97 | 1.90 | -2.07 | 3.23 | 0.84 | -0.02 | -- | -- | -- |
| $Mn_2Co_{1.08}Ga_{0.92}$ | 2.54 | 2.41 | -2.09 | 3.2 | 0.88 | -0.02 | 3.25 | 1.16 | -- |
| $Mn_2Co_{1.16}Ga_{0.84}$ | 2.75 | 2.92 | -2.14 | 3.18 | 0.94 | -0.03 | 3.2 | 1.18 | -- |
| $Mn_2Co_{1.32}Ga_{0.68}$ | 3.95 | 3.92 | -2.24 | 3.12 | 1.04 | -0.05 | 3.13 | 1.23 | -- |
| $Mn_2CoCu_{0.08}Ga_{0.92}$ | 2.4 | 2.32 | -2.1 | 3.23 | 0.85 | -0.02 | 3.25 | 0.03 | -- |
| $Mn_2CoCu_{0.16}Ga_{0.84}$ | 2.65 | 2.76 | -2.16 | 3.24 | 0.89 | -0.02 | 3.22 | 0.04 | -- |
| $Mn_2CoCu_{0.32}Ga_{0.68}$ | 3.37 | 3.67 | -2.28 | 3.24 | 1.00 | -0.03 | 3.21 | 0.05 | -- |
| $Mn_2CoV_{0.08}Ga_{0.92}$ | 2.1 | 2.02 | -2.01 | 3.19 | 0.85 | -0.02 | -- | -- | 0.95 |
| $Mn_2CoV_{0.16}Ga_{0.84}$ | 2.24 | 2.15 | -1.96 | 3.16 | 0.87 | -0.02 | -- | -- | 0.97 |
| $Mn_2CoTi_{0.08}Ga_{0.92}$ | 1.91 | 1.92 | -2.04 | 3.22 | 0.83 | -0.02 | -- | -- | 0.03 |
| $Mn_2CoTi_{0.16}Ga_{0.84}$ | 1.94 | 1.93 | -1.99 | 3.18 | 0.82 | -0.02 | -- | -- | 0.02 |

Table I collects the experimentally measured and calculated molecular moments, as well as the moments of each atom in $Mn_2CoM_xGa_{1-x}$ alloys at the corresponding atomic sites. We conclude from the calculated results that Mn($A$) carries a smaller moment of $-2.07\mu_B$ than that of Mn($B$) of $3.23\mu_B$, which is because of the covalent effect from the main group element of Ga.[29, 34] Their moments antiferromagnetically align, due to the nearest neighboring distance. The moment of Co($C$) has a value of $0.82\sim1.0\mu_B$ and ferromagnetically couples with Mn($B$). Thus Mn($A$), Mn($B$) and Co($C$) compose a native ferrimagnetic matrix in the parent alloy system.[7, 9, 35]

When Co substitutes for Ga, some Mn($A$) atoms become Mn($D$) due to the occupation rule and their moment value increases from about 2.07 to $3.25\mu_B$. Meanwhile, the sign changes from negative to positive (see Table I). These changes result in the Mn($A$) atom with a relatively small moment being removed from the antiferromagnetic coupling and added into the ferromagnetic coupling as Mn(*D*) with a relatively large moment.[9] Thus, a strong ferromagnetic structure, Co(*A*)-Mn(*B/D*)-Co(*C*), is achieved as the nearest neighbors in the ferrimagnetic matrix. This simultaneously enhances the ferromagnetic coupling and decreases the antiferromagnetic coupling, leading to the dramatical enhancement of the magnetization of $\Delta m_{Co} = 6.18\mu_B$.

For the Cu doping, the Mn(*D*) can also be generated as the Co doping case, so a different ferromagnetic structure, Mn(*A*)-Co(*C*)-Mn(*D*) is achieved, which leads to the $\Delta m_{Cu}$ of $5.24\mu_B$. This is the reason why the non-magnetic Cu can enhance the magnetization in $Mn_2CoGa$ system. But the doped V atom occupies the *D* site, because its valence electron number is less than that of Mn. This atomic configuration forms a new ferromagnetic structure of Mn(*B*)-Co(*C*)-V(*D*) due to the strong exchange interaction effect of Co.[36, 37] One may find that, from Table I, the increment of $\Delta m_V$ ($1.57\mu_B$) mainly comes from the V(*D*) moment of about $1\mu_B$. The previous works reported that the moment of the V can have a value of about $1.1\mu_B$, but antiferromagnetically couples with Mn in Heusler alloys $Mn_2VAl$.[35, 38] In a ferromagnetically coupling case, the moment of the V becomes smaller of about $0.7\mu_B$ in $Co_2VSi$[39]. Therefore, the value of about $1.0\mu_B$ is the largest ferromagnetic moment for the V in Heusler alloys known in the references.

There is not apparent change of the magnetization for the Ti doping, because it is not a magnetic element and does not result in a magnetic structure change either. Calculations were also performed by assuming the Co and Cu occupy the *D* sites, but it produces a strange negative moment for Co and an abnormally low total moment, far from the experimental value, indicating that the practical atomic configuration really obeys the occupation rule.

Similar to the case in $Mn_2CoGa$, the increase of magnetization has also been observed in $Mn_2CoAl$ alloy when substituting Co for Al, although the *B-D* site disordered occupation occurs as shown in Fig. 3. This is due to that the Co(*A*)-Mn(*B/D*)-Co(*C*) ferromagnetic structure can still be achieved in such partly B2 type disordered situation.

**D. Exchange interaction and electronic structure calculation**

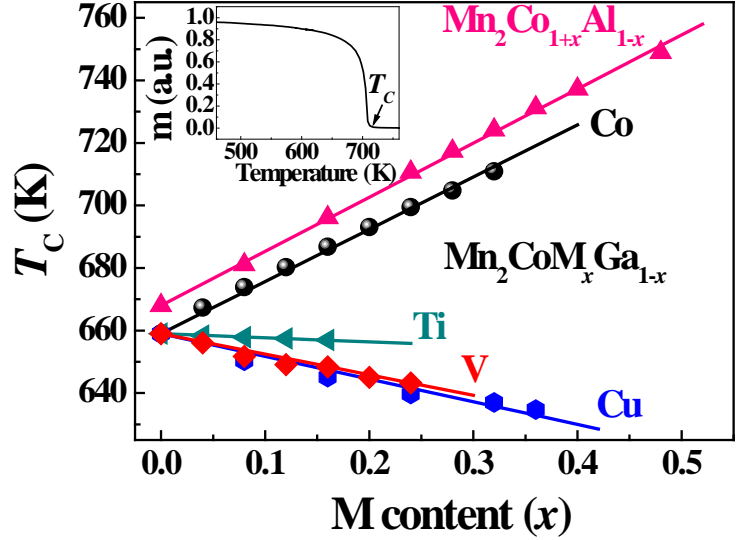

FIG. 6. (Color online) Curie temperature ($T_C$) as a function of doping concentration $x$ in systems of $Mn_2CoM_xGa_{1-x}$ (M = Co, Cu, V and Ti) and $Mn_2Co_{1+x}Al_{1-x}$. The straight lines are the results of linear fitting.

Figure.6 shows the composition dependence of Curie temperature ($T_C$) for the various dopants. When doping Co in $Mn_2CoAl$ and $Mn_2CoGa$, the $T_C$ increases with the almost same rate. Naturally, it should be attributed to the strong exchange interaction[40] in the Co-Mn-Co ferromagnetic structure. Turning to the case for other elements doping, however, their $T_C$ decrease with different rates. The $T_C$ just only has a slight reduction for doping Ti, corresponding to the unconspicuous change of the moment in $Mn_2CoTi_xGa_{1-x}$. Comparing with the molecular moment shown in Fig. 5, one may note that the $T_C$ has a relatively large decrease for doping Cu and V, contrasting to their apparent increase of molecular moments. Because the doped Cu is a non-magnetic element, it does not contribute the magnetic exchange interaction for the system. Furthermore, the Cu doping cause the Mn(*D*) occupation, which create a local ferromagnetic structure. Although doped V atoms have ferromagnetic moment aligned with Co and Mn($B$), the exchange interaction energy contributed by V is not enough to cover the energy depleted by achieving a Mn-Co-V ferromagnetic structure. In both cases, the achievement of ferromagnetic structure is at cost of the depletion of the magnetic exchange interaction energy. Therefore, in contrast to the Co doping, the $T_C$ in the materials doped by Cu and V is decreased.

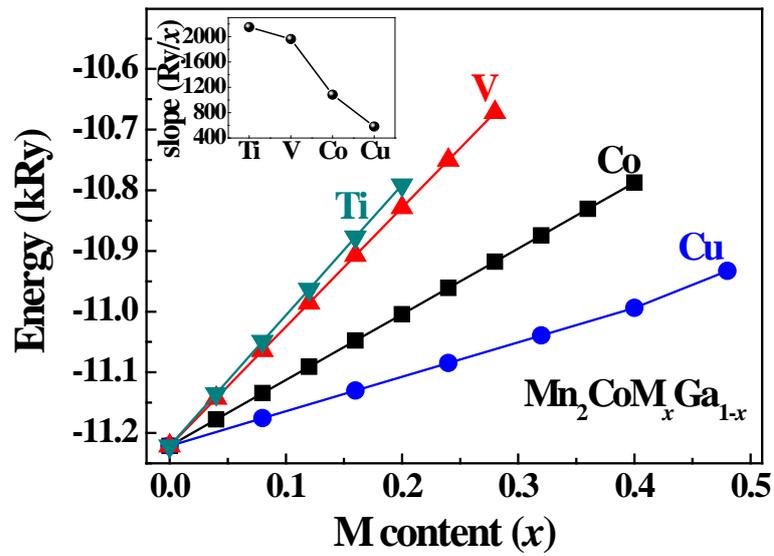

FIG. 7. (Color online) The M content dependence of the total energy in $Mn_2CoM_xGa_{1-x}$ (M= Co Cu, V, and Ti). Inset is the slope for total energy increments for different dopants.

Figure 7 presents the calculation results about the total energy with the variety of doping concentration. It can be found that all dopants increase the total energy of the systems. This reflects a tolerance for the substitution: the newly adding element can cause instability of systems, which should be the reason for the second phase appearing at a certain doping concentration, as shown in Fig. 2 (c). From Fig. 7, we can also find the energy increment diminishes as the atomic number increase. It is because that the covalent effect becomes stronger when the atom has more valence electrons and it enhances the stability of the alloy system.

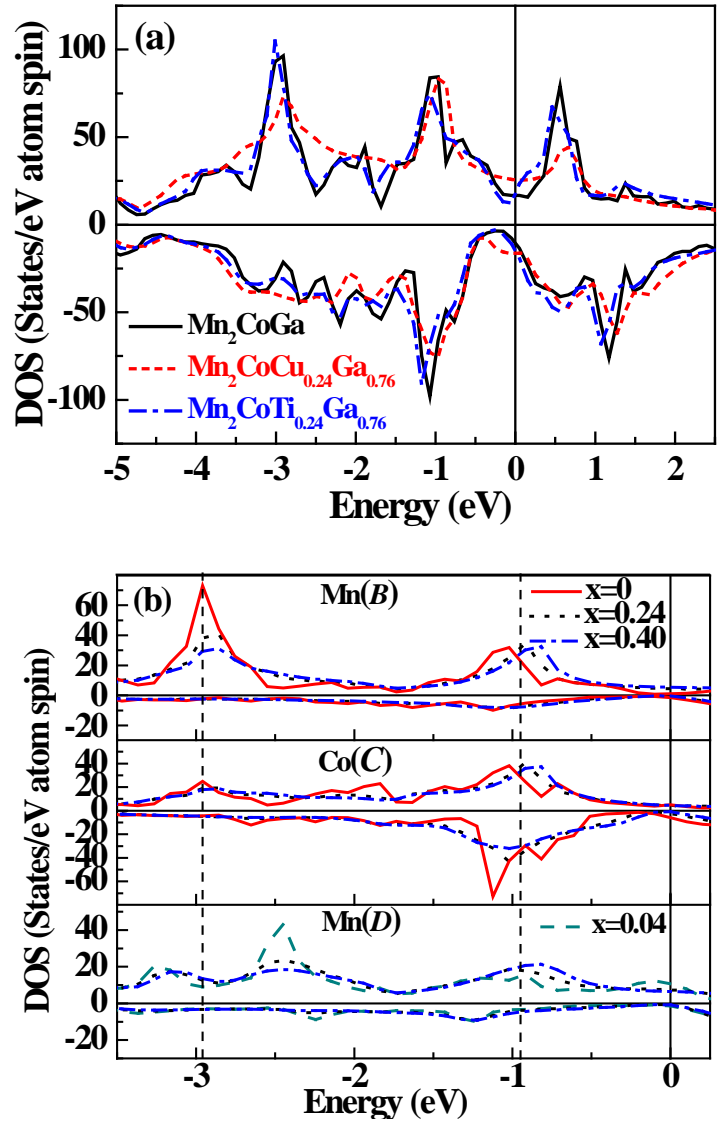

FIG. 8. (Color online) Calculated spin-projected DOS plots for the $Mn_2CoM_xGa_{1-x}$ systems: the total DOS for the Cu and Ti dopants (a) and partial DOS for $d$-electrons as M = Co (b). The upper halves display the spin-up states.

Figure 8 shows the calculated total DOS and partial $d$-electron DOS for Cu, Ti and Co doped systems. The total DOS ranging from -5.0 to +2.5 eV mainly consist of $d$-electrons of Co and Mn atoms. As shown in Fig. 8(a), the doping of non-magnetic elements, Cu and Ti, causes the different dispersed occupied states. The apparent widening suggests a strong $d$-$d$ hybridization between Co and

Mn atoms in Cu doped system.[41] It results the dramatically increase of the molecular moment (Fig. 5). In contrast, the DOS of Ti doped system does not show the significant change. This implies there is not variation for exchange interaction in this system. In the case of Co doping as shown in Fig. 8(b), the hybridized peaks between Mn(*B*) and Co(*C*) atoms, around -0.95 and -2.96 eV in spin-up direction (marked with dash lines), are widened as Co doping, showing an ordinary hybridization enhancement.[7, 9, 13] But the up-spin of *d*-electrons of Mn(*D*) shows an apparent shift to high energy and an increase of intensity as increasing Co content, which also indicates an enhanced hybridization. This corresponds to the achievement of a ferromagnetic structure in the nearest neighboring distance. In the present work, the electronic structures for V doped system and Co doped $Mn_2CoAl$ system have also been calculated. The electronic structure for V doping is almost the same with that for Ti doping and the Co doped in $Mn_2CoAl$ is exactly similar with that in $Mn_2CoGa$.

## V. CONCLUSION

The doped Heusler alloys of $Mn_2CoGa$ and $Mn_2CoAl$ have been investigated about structure and magnetic properties by experiments and calculations. In order to study the magnetic structure, some magnetic or non-magnetic transition metals, Co, Cu, V, and Ti have been used to substitute the main group elements of Ga and Al in the alloy systems. It has been found that there are three kinds of local ferromagnetic structures, the Co-Mn-Co, Mn-Co-Mn and Mn-Co-V, embedded in the native ferrimagnetic matrix. They increase the magnetization of the parent alloy with the different increments. Therefore, the phenomena for non-magnetic Cu increasing magnetization and the large V moment of about $1.0\mu_B$ have been observed. The experimental results also indicate that the newly added Co enhances the exchange interaction and increases the $T_C$, while the decreased $T_C$ reflects a depletion of the exchange interaction energy for achieving the ferromagnetic coupling in Mn-Co-Mn and Mn-Co-V structures. The strong *d-d* hybridization has been verified by our calculation of electronic structure, when the ferromagnetic structures are achieved. The covalent effect of the main group elements plays an important role in our samples. From it, the occupation rule arranges the doped atoms in occupy *A* or *D* sites, which determines if the local ferromagnetic structure can be generated. The covalent effect also affects the system stability and doping tolerance in the studied systems.

**ACKNOWLEDGMENTS**

This work is supported by the National Natural Science Foundation of China in Grant No. 51031004 and 11174352 and National Basic Research Program of China (973 Program, 2012CB619405).

References:

[1] P. J. Webster, Contemporary Physics **10** (6), 559-577 (1969).

[2] K. H. J. Buschow and P. G. van Engen, Journal of Magnetism and Magnetic Materials **25** (1), 90-96 (1981).

[3] R. A. de Groot, F. M. Mueller, P. G. vanEngen and K. H. J. Buschow, Physical Review Letters **50** (25), 2024-2027 (1983).

[4] I. Galanakis, P. H. Dederichs and N. Papanikolaou, Physical Review B **66** (17), 174429 (2002).

[5] R. Kainuma, Y. Imano, W. Ito, Y. Sutou, H. Morito, S. Okamoto, O. Kitakami, K. Oikawa, A. Fujita, T. Kanomata and K. Ishida, Nature **439** (7079), 957-960 (2006).

[6] A. Planes, L. Mañosa, X. Moya, T. Krenke, M. Acet and E. F. Wassermann, Journal of Magnetism and Magnetic Materials **310** (2, Part 3), 2767-2769 (2007).

[7] G. D. Liu, X. F. Dai, H. Y. Liu, J. L. Chen, Y. X. Li, G. Xiao and G. H. Wu, Physical Review B **77** (1), 014424 (2008).

[8] L. Feng, L. Ma, Z. Y. Zhu, W. Zhu, E. K. Liu, J. L. Chen, G. H. Wu, F. B. Meng, H. Y. Liu, H. Z. Luo and Y. X. Li, Journal of

Applied Physics **107** (1), 013913 (2010).

[9]L. Ma, W. H. Wang, C. M. Zhen, D. L. Hou, X. D. Tang, E. K. Liu and G. H. Wu, Physical Review B **84** (22), 224404 (2011).

[10]K. Le Dang, P. Veillet and I. A. Campbell, Journal of Physics F: Metal Physics **8** (8), 1811 (1978).

[11]K. Özdogan and et al., Journal of Physics: Condensed Matter **18** (10), 2905 (2006).

[12]K. Ramesh Kumar, J. Arout Chelvane, G. Markandeyulu, S. K. Malik and N. Harish Kumar, Solid State Communications **150** (1-2), 70-73 (2010).

[13]G. D. Liu, X. F. Dai, S. Y. Yu, Z. Y. Zhu, J. L. Chen, G. H. Wu, H. Zhu and J. Q. Xiao, Physical Review B **74** (5), 054435 (2006).

[14]K. R. A. Ziebeck and P. J. Webster, Journal of Physics and Chemistry of Solids **35** (1), 1-7 (1974).

[15]Z. H. Liu, M. Zhang, Y. T. Cui, Y. Q. Zhou, W. H. Wang, G. H. Wu, X. X. Zhang and G. Xiao, Applied Physics Letters **82** (3), 424-426 (2003).

[16]H. Katayama, K. Terakura and J. Kanamori, Solid State Communications **29** (5), 431-434 (1979).

[17]S. Blügel, H. Akai, R. Zeller and P. H. Dederichs, Physical Review B **35** (7), 3271-3283 (1987).

[18]S. Kaprzyk and A. Bansil, Physical Review B **42** (12), 7358 (1990).

[19]H. Akai, Hyperfine Interactions **68** (1), 3-14 (1992).

[20]M. Ogura and H. Akai, Applied Physics Letters **91** (25), 253118-253113 (2007).

[21]T. J. Burch, J. I. Budnick, V. A. Niculescu, K. Raj and T. Litrenta, Physical Review B **24** (7), 3866-3883 (1981).

[22]T. J. Burch and T. Litrenta, Physical Review Letters **33** (7), 421-424 (1974).

[23]J. Enkovaara, O. Heczko, A. Ayuela and R. M. Nieminen, Physical Review B **67** (21), 212405 (2003).

[24]A. Bansil, S. Kaprzyk, P. E. Mijnarends and J. Toboła, Physical Review B **60** (19), 13396 (1999).

[25]K. Hem Chandra and et al., Journal of Physics D: Applied Physics **39** (5), 776 (2006).

[26]L. Hongzhi and et al., Journal of Physics D: Applied Physics **41** (5), 055010 (2008).

[27]A. M. Bratkovsky, S. N. Rashkeev and G. Wendin, Physical Review B **48** (9), 6260-6270 (1993).

[28]H.-M. Hong, Y.-J. Kang, J. Kang, E. C. Lee, Y. H. Kim and K. J. Chang, Physical Review B **72** (14), 144408 (2005).

[29]J. Kübler, A. R. Williams and C. B. Sommers, Physical Review B **28** (4), 1745-1755 (1983).

[30]Z. Q. Feng, H. Z. Luo, Y. X. Wang, Y. X. Li, W. Zhu, G. H. Wu and F. B. Meng, Phys. Status Solidi A **207** (6), 1481–1484 (2010).

[31]R. Saha, V. Srinivas and T. V. Chandrasekhar Rao, Physical Review B **79** (17), 174423 (2009).

[32]P. J. Webster, Journal of Physics and Chemistry of Solids **32** (6), 1221-1231 (1971).

[33]in *WebElements* (http://www.webelements.com/).

[34]A. R. Williams, R. Zeller, V. L. Moruzzi, J. C. D. Gelatt and J. Kubler, Journal of Applied Physics **52** (3), 2067-2069 (1981).

[35]M. Markus and et al., Journal of Physics D: Applied Physics **44** (21), 215003 (2011).

[36]M. B. Stearns, Journal of Applied Physics **50** (B3), 2060-2062 (1979).

[37]M. B. Stearns, Journal of Applied Physics **49** (3), 2165-2167 (1978).

[38]I. Galanakis, K. Ozdogan, E. Sasioglu and B. Aktas, Physical Review B **75** (9), 092407 (2007).

[39]H. C. Kandpal, C. Felser and G. H. Fecher, Journal of Magnetism and Magnetic Materials **310** (2, Part 2), 1626-1628 (2007).

[40]M. Markus and et al., Journal of Physics: Condensed Matter **23** (11), 116005 (2011).

[41]S. Picozzi, A. Continenza and A. J. Freeman, Physical Review B **66** (9), 094421 (2002).